\documentclass[fleqn,twoside]{article}
\usepackage{espcrc2}

\usepackage{graphicx}
\usepackage[figuresright]{rotating}

\def\beq{\begin{equation}}
\def\eeq{\end{equation}}
\def\be{\begin{eqnarray}}
\def\ee{\end{eqnarray}}

\def\gsim{\buildrel > \over {_{\sim}}}

\newcommand{\AmS}{{\protect\the\textfont2
  A\kern-.1667em\lower.5ex\hbox{M}\kern-.125emS}}

\title{Final state interactions in the electroweak nuclear response}

\author{Omar Benhar \\ ~~~ \\
INFN and Department of Physics, Universit\`a ``La Sapienza'' \\
        Piazzale Aldo Moro, 2. I-00185 Roma, Italy}
       
\begin{document}

\begin{abstract}
I review the description of the electroweak nuclear response at large momentum 
transfer within nonrelativistic many-body theory. Special consideration is given 
to the effects of final state interactions, which are known to be large in
both inclusive and semi-inclusive processes. The results of theoretical 
calculations of electron-nucleus scattering observables are compared to the data,  
and the generalization to charged current neutrino-nucleus interactions is 
discussed.
\vspace{1pc}
\end{abstract}

\maketitle

\section{INTRODUCTION}

Over the past few years the rapid development of neutrino physics, 
leading to significant improvements in the experimental accuracy,  
has triggered a burst of studies aimed at reducing the systematic
uncertainty associated with the treatment of nuclear effects. The
main results of these activities are discussed in the Proceedings
of the previous meetings in the  NUINT series \cite{NUINT01,NUINT04}.

It has soon been realized that much of the information needed to 
understand nuclear effects at quantitative level can be extracted 
from the large body of electron-nucleus scattering data \cite{Elba}, 
and that the theoretical techniques developed to describe
the nuclear response to electromagnetic probes can be readily 
generalized to obtain accurate
predictions of neutrino-nucleus scattering observables.

In this paper I review the approach based on
nonrelativistic nuclear many-body theory (NMBT), that allows one to 
consistently include the effects of dynamical nucleon-nucleon (NN) 
correlations in both the initial and final states.
The impulse approximation (IA) scheme, in which the
cross section is written in terms of the nuclear spectral 
function $P(p)$, describing the momentum and energy distribution
of nucleons in the target nucleus,
is outlined in Section 2. Section 3 is devoted to the analysis of 
final state interactions (FSI), whose effects are known to be large, 
while in Section 4 the results of theoretical calculations are discussed
and compared to electron-nucleus scattering data. Finally, in Sections 5 
I summarize the main results and state the conclusions.

\section{THE IMPULSE APPROXIMATION}

Let us consider the process
\beq
\ell + A \rightarrow \ell^\prime + X\ ,
\eeq
in which $\ell$ and $\ell^\prime$ denote either a charged lepton or 
a neutrino, and the final state of the target nucleus is unobserved.
The corresponding differential cross section can be written in the form
\beq
\nonumber
\frac{ d\sigma }{ d\Omega_{\ell^\prime} dE_{\ell^\prime} } \propto
L_{\mu\nu} W_A^{\mu\nu} \ ,
\label{sigma:IA}
\eeq
where $\Omega_{\ell^\prime}$ and $E_{\ell^\prime}$ are the scattering angle and 
energy of the outgoing lepton, respectively. The tensor $L_{\mu\nu}$ is totally 
specified by kinematics, whereas the definition of the target response 
tensor 
\beq
W_A^{\mu\nu} = \sum_X \langle 0 | {J_A^\mu}^\dagger | X \rangle
      \langle X | J_A^\nu | 0 \rangle \delta^{(4)}(p_0 + q - p_X)\ ,
\label{tens:resp}
\eeq
involves the hadronic initial and final states $| 0 \rangle$ and $| X \rangle$, 
 carrying four-momenta $p_0$ and $p_X$, respectively, as well as the nuclear 
electroweak current operator ${J_A^\mu}$. 

Calculations of $W_A^{\mu\nu}$ of Eq.~(\ref{tens:resp}) at 
moderate momentum transfer $( {\bf |q|} < 0.5\, {\rm GeV})$
can be carried out within NMBT, using
nonrelativistic wave functions and expanding the current operator in powers 
of ${\bf |q|}/m$, where $m$ is the nucleon mass \cite{Carlson98}.

At higher momentum transfer, corresponding to beam energies 
larger than $\sim 1$ \ GeV, describing the final states $|X\rangle$
in terms of nonrelativistic nucleons is no longer possible.
Due to the prohibitive difficulties involved in a fully consistent treatment of the
relativistic nuclear many-body problem, calculations of $W_A^{\mu\nu}$ in this 
regime require a set of simplifying assumptions, allowing one to take into account 
the relativistic motion of
final state particles carrying momenta $\sim {\bf q}$, as well as the 
possible occurrence of inelastic processes leading to the appearance of hadrons 
other than protons and neutrons.

The main assumption underlying IA is that, as
the space resolution of a probe delivering momentum ${\bf q}$ 
is $\sim 1/|{\bf q}|$, at large enough $|{\bf q}|$
the target is seen by the probe as a collection of individual nucleons.
Hence, in the IA regime, the scattering process off a nuclear target reduces to the
incoherent sum of elementary processes involving only one nucleon\footnote{
Coherent contributions, not taken into account in the impulse approximation,
play a role even at large $|{\bf q}|$ for values of the Bjorken scaling variable
$x < 0.2$, corresponding to very large lepton energy loss.
However, they are not relevant to the kinematical regime discussed in this paper.
}. 

\begin{figure}[hbt]
\begin{center}
\vspace*{-0.50in}
\includegraphics[scale=0.80]{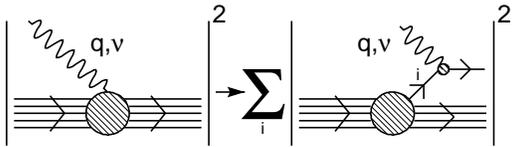}
\end{center}
\vspace*{-.5in}
\caption{\small
Pictorial representation of the PWIA scheme, in which
the nuclear cross section is replaced by the incoherent sum of
cross sections describing scattering off individual nucleons, the
recoiling (A-1)-particle system acting as a spectator.
}
\label{fig:2}
\vspace*{-.10in}
\end{figure}

The simplest implementation of IA, referred to as Plane Wave Impulse Approximation
(PWIA) is based on the further assumption that the effects of FSI 
between the hit nucleon and the (A-1)-nucleon spectator system be negligible. The 
resulting picture of the scattering process is schematically illustrated in 
Fig. \ref{fig:2}.

Within PWIA the target response tensor of Eq.~(\ref{tens:resp}) reduces to
\beq
W^{\mu\nu}_A(q) = \int d^4p\  P(p)\  {\widetilde W}^{\mu\nu}(p,q) \ ,
\label{tens:IA}
\eeq
where $q\equiv(\nu,{\bf q})$ is the four-momentum transfer. The nuclear spectral 
function $P(p)$, with $p\equiv(M_A-E_{A-1},{\bf p})$, yields the probability 
of removing a nucleon with momentum ${\bf p}$ from the target ground state 
leaving the residual system with energy $E_{A-1}$ \cite{PKE,LDA}, whereas the tensor 
${\widetilde W}^{\mu\nu}$ describes the electroweak interactions of a 
{\it bound} nucleon. Within IA binding effects can be easily taken into account 
setting  \cite{defo}
\beq
{\widetilde W}^{\mu\nu}(p,q) = W^{\mu\nu}(p,{\widetilde q}) ,
\label{defo:1}
\eeq
where $W^{\mu\nu}$ is the tensor associated with a {\it free} nucleon, that can be 
expressed in terms of the measured structure functions, and 
${\widetilde q}\equiv({\widetilde \nu},{\bf q})$, with 
\beq
{\widetilde \nu} = \nu + M_A - E_{A-1} -
\sqrt{|{\bf p}|^2 + m^2}\ .
\label{defo:2}
\eeq
According to Eqs. (\ref{defo:1})-(\ref{defo:2}) 
a fraction $(\nu - {\widetilde \nu})/\nu$ of the lepton energy loss is 
spent to put the struck nucleon on the mass shell, and the elementary
scattering process is described as if it took place in free space
with energy transfer ${\widetilde \nu}<\nu$.

While being fully justified on physics grounds, as part of the lepton energy 
loss does go into excitation energy of the spectator system, the replacement
of $\nu$ with ${\widetilde \nu}$ poses a non-trivial
conceptual problem, in that it leads to a violation of vector current
conservation. However, this issue turns out to be only marginally relevant,  
since the non gauge invariant contributions can 
be shown to vanish in the $|{\bf q}| \rightarrow \infty$ limit. 

\section{FINAL STATE INTERACTIONS}

The occurrence of strong FSI in quasi-elastic electron-nuclus scattering has 
long been experimentally established. One of the most striking evidences is the 
loss of flux of outgoing particles observed in electron induced proton knock-out
experiments \cite{Garino92,O'Neill95,Abbott98,Garrow02,Rohe05}. The suppression of the 
measured nuclear transparencies with respect to the PWIA limit turns out to be 
as strong as 20-40 \% in Carbon and 50-70 \% in Gold.

A theoretical description of FSI based on NMBT and a generalization of Glauber 
theory of high energy proton scattering \cite{Glauber59} has been proposed 
in the early 90's \cite{Benhar91}. This approach, generally
referred to as Correlated Glauber Approximation (CGA), rests on the premises that
i) the struck nucleon moves along a straight trajectory with constant velocity
(eikonal approximation), and ii) the spectator nucleons are seen by the
struck particle as a collection of fixed scattering centers
(frozen approximation).

Under the above assumptions the propagator, describing the struck nucleon at 
time $t$ after 
the electroweak interaction, can be written in the factorized form
\cite{Petraki03}
\beq
U_{{\bf p}+{\bf q}}(t) = U^0_{{\bf p}+{\bf q}}(t)
{\bar U}^{FSI}_{{\bf p}+{\bf q}}(t)\ ,
\eeq
where $U^0_{{\bf p}+{\bf q}}(t)$ is the free space propagator, while FSI
effects are described by the quantity
\beq
{\bar U}^{FSI}_{{\bf p}+{\bf q}}(t) = 
\langle 0 | U^{FSI}_{{\bf p}+{\bf q}}({\bf r}_1,{\widetilde R};t) | 0 \rangle \ .
\label{eik:prop0}
\eeq
Here ${\bf r}_1$ and ${\widetilde R}\equiv({\bf r}_2 \ldots {\bf r}_A)$ 
specify the positions of the struck particle and the spectators, respectively, 
$\langle 0 | \ldots | 0 \rangle $ denotes the expectation value in the target ground state 
and
\beq
U^{FSI}_{{\bf p}+{\bf q}}({\bf r}_1,{\widetilde R};t) =  
{\rm e}^{-i \sum_{j} \int_0^t dt^\prime
\Gamma_{{\bf p}+{\bf q}}(|{\bf r}_{1j} + {\bf v}t^\prime |) } \ .
\label{eik:prop}
\eeq

In Eq.~(\ref{eik:prop}), ${\bf r}_{1j}={\bf r}_{1}-{\bf r}_{j}$ $(j=2,\ldots A)$ 
and $\Gamma_{{\bf p}+{\bf q}}(|{\bf r}|)$ is the coordinate-space
t-matrix, simply related to the measured nucleon-nucleon (NN) scattering 
amplitude at incident momentum ${\bf p}+{\bf q}$.
At large $|{\bf q}|$, ${\bf p}+{\bf q} \approx {\bf q}$ and the eikonal
propagator of Eq. (\ref{eik:prop0}) becomes a function
of $t$ and the momentum transfer only.

The quantity
\beq
P_{\bf q}(t) = 
\langle 0 | |U^{FSI}_{{\bf q}}({\bf r}_1,{\widetilde R};t) |^2 | 0 \rangle \
\label{T:1}
\eeq
measures the probability that the struck nucleon do not undergo rescattering
processes during a time $t$ after the electroweak interaction. In absence of FSI,
i.e. for vanishing $\Gamma_{{\bf q}}$, $ P_{\bf q}(t) \equiv 1$.
Note that $P(t)$ is trivially related to the nuclear
transparency $T_{{\bf q}}$, measured in coincidence $(e,e^\prime p)$
experiments \cite{Garino92,O'Neill95,Abbott98,Garrow02,Rohe05}, through
\beq
T_{A} = \lim_{t \rightarrow \infty} P_{{\bf q}}(t)
\label{T:2}
\eeq

It is very important to realize that, as shown by 
Eqs.~(\ref{eik:prop0})-(\ref{T:1}), the probability that a
rescattering process occur is not simply dictated by
the nuclear density distribution $\rho_A({\bf r}_j)$, yielding the probability
of finding a spectator
at position ${\bf r}_j$. It depends upon the {\it joint} probability of finding the
 struck particle at position ${\bf r}_1$ {\it and} a spectator at position
${\bf r}_j$, that can be written in the form
\beq
\rho^{(2)}({\bf r}_1,{\bf r}_j) =
\rho_A({\bf r}_1)\rho_A({\bf r}_j) g({\bf r}_1,{\bf r}_j) \ .
\label{def:rho2}
\eeq

Due to the strongly repulsive nature of nuclear interactions at short range,
$\rho^{(2)}({\bf r}_1,{\bf r}_j)$ is largely affected by NN correlations,
 whose effect is described by the correlation function $g({\bf r}_1,{\bf r}_j)$.
The results of numerical calculations carried out within NMBT yield 
$g({\bf r}_1,{\bf r}_j) \ll 1$ at $|{\bf r}_{1j}| < 1$ fm. 

\begin{figure}[hbt]
\begin{center}
\includegraphics[scale=0.55]{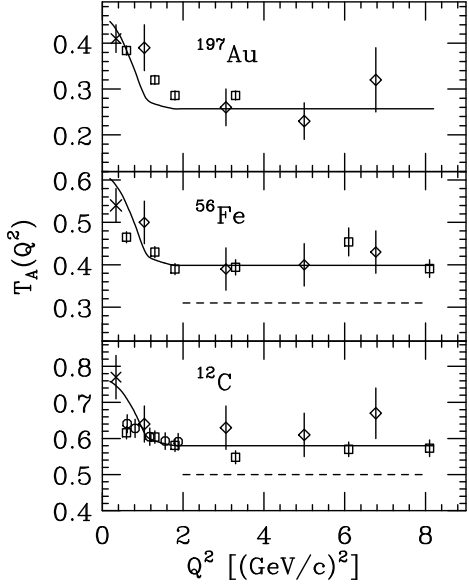}
\end{center}
\vspace*{-.5in}
\caption{\small
$Q^2$-dependence of the transparency of Carbon, Iron and Gold,
calculated within the approach of Ref.\protect\cite{Benhar91}. The data points
are taken from
Refs.\protect\cite{Garino92} (crosses), \cite{O'Neill95} (diamonds),
\cite{Abbott98,Garrow02} (squares) and \cite{Rohe05} (circles). 
The dashed lines have been obtained neglecting the effect of NN 
correlations, i.e. setting $g({\bf r}_1,{\bf r}_j) \equiv 1$
in Eq. (\protect\ref{def:rho2}). Note that in absence of FSI $T_A(Q^2) \equiv 1$.
}
\label{transp}
\vspace*{-.10in}
\end{figure}

The results displayed in Fig. \ref{transp} show that both the magnitude
and the $A$- and $Q^2$-dependence of the transparencies of Carbon, Iron and Gold
obtained from the approach of Ref. \cite{Benhar91} are in good agreement
with the experimental data. Note that in absence of FSI $T_A(Q^2) \equiv 1$. 

The calculated nuclear transparencies turn out to be strongly affected by NN 
correlations. Neglecting their effects by setting $g({\bf r}_1,{\bf r}_j) \equiv 1$
in Eq. (\ref{def:rho2}), one obtains $T_A\approx$~0.5 and 0.3 for Carbon and Iron, 
respectively,  at $Q^2>2$. Figure \ref{transp} shows that these values are utterly 
incompatible with the data. 

Being only sensitive to rescattering processes taking place within a distance
$\sim 1/|{\bf q}|$ of the electroweak vertex, the inclusive cross section at
high momentum transfer is much less affected by FSI than the $(e,e^\prime p)$
cross section. However, FSI effects are appreciable, and become dominant 
in the low $\nu$ tail, where PWIA calculations largely underestimate 
electron-nucleus scattering data.

In inclusive processes FSI have two effects: i) an energy shift of the cross section,
due to the fact that the struck nucleon moves in the average potential generated by the
spectator particles and ii) a redistribution of the strength, leading to the quenching
of the quasielastic peak and the enhancement of the tails, as a consequence of
the occurrence of NN scattering processes coupling the one particle-one hole 
final state to more complex n-particle n-hole configurations.

According to Ref. \cite{Benhar91}, in presence of FSI the inclusive cross section 
can be expressed in terms of the PWIA result through 
\beq
\frac{d\sigma}{d\Omega_{\ell^\prime} d\nu} = \int d\nu^\prime 
\left( \frac{ d\sigma }{ d\Omega_{\ell^\prime} d\nu^\prime } \right)_{PWIA} 
f_{{\bf q}}(\nu - \nu^\prime) ,
\label{sigma:FSI}
\eeq
the folding function $f_{{\bf q}}(\nu)$ being defined as
\be
\nonumber
f_{{\bf q}}(\nu) & = & \delta(\nu) \sqrt{ T_{A} } \\
& + &
\int \frac{dt}{2 \pi}\ {\rm e}^{i \nu t}
\left[ {\bar U}^{FSI}_{{\bf q}}(t) - \sqrt{ T_{A} } \right]\ .
\label{ff}
\ee
The above equations clearly show that the strength of FSI is measured by
both $T_{A}$ and the width of the folding function. In absence of FSI, 
${\bar U}^{FSI}_{{\bf q}}(t) \equiv 1$, implying in turn
$T_{A}=1$ and $f_{{\bf q}}(\nu) \rightarrow \delta(\nu)$.

\section{RESULTS}

The approach described in the previous Sections has been employed to carry out
calculations of the inclusive cross sections for both electron-nucleus and 
charged current neutrino-nucleus processes.

\begin{figure}[hbt]
\begin{center}
\vspace*{-.10in}
\includegraphics[scale=0.40]{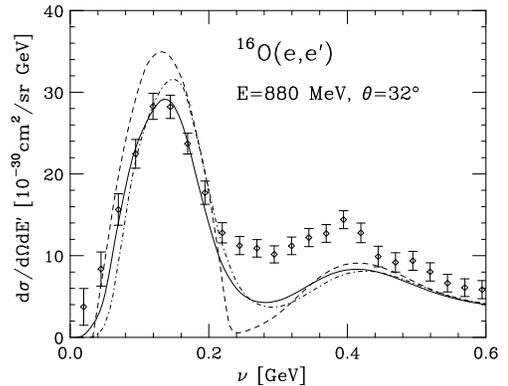}
\end{center}
\vspace*{-.5in}
\caption{\small
Cross section of the process $^{16}O(e,e^\prime)$
at beam energy 880 MeV and electron scattering angle 32$^\circ$.
Solid line: full calculation.  Dot-dash line:
PWIA calculation, carried out neglecting FSI effects.
Dashed line: FG model with $p_F = 225$ MeV and $\epsilon = 25$ MeV. The experimental
data are from Ref.\protect\cite{LNF}.
}
\label{fig:1}
\vspace*{-.10in}
\end{figure}

In Fig. \ref{fig:1} the cross section of the process 
$e~+~^{16}~O~\rightarrow~e^\prime~+~X$, obtained from
Eq. (\ref{sigma:FSI}) \cite{Benhar05}, is 
compared to the data of Ref. \cite{LNF}.
The results of theoretical calculation, {\it involving no adjustable parameters}, 
provide a very accurate description of the measured cross sections in the region 
of the quasi-elastic peak. The effect of FSI, leading to a shift and a quenching 
of the peak, is clearly visible.
For reference, the figure also shows the results of the Fermi gas (FG) model,
corresponding to Fermi momentum $p_F = 225$ MeV and nucleon removal
energy $\epsilon = 25$ MeV, which appears to largely overestimate the data.
The failure of the theoretical calculations to reproduce the measured cross section in
the region of the $\Delta$-production peak is likely to be ascribable to deficiencies 
in the description of the elementary electron-nucleon cross section \cite{Benhar05}.

\begin{figure}[hbt]
\begin{center}
\vspace*{-.10in}
\includegraphics[scale=0.40]{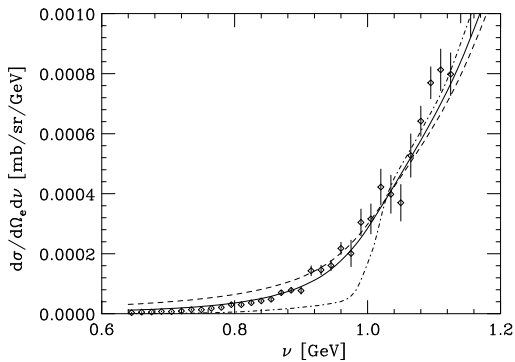}
\end{center}
\vspace*{-.5in}
\caption{\small
Comparison between the results of the approach of Ref. \protect\cite{Benhar91}
and the extrapolated nuclear matter cross section of Ref. \protect\cite{NMXS} at
$E_e = 3.6$ GeV and $\theta_{e^\prime}=30^\circ$. Dash-dot line: PWIA calculation.
Solid line: full calculation, including FSI. The dashed line has been obtained
neglecting the effects of NN correlations in the calculation of FSI effects,
i.e. setting $g({\bf r}_1,{\bf r}_j) \equiv 1$
in Eq. (\protect\ref{def:rho2}).
}
\label{fig:3}
\vspace*{-.10in}
\end{figure}

\begin{figure}[hbt]
\begin{center}
\includegraphics[scale=0.40]{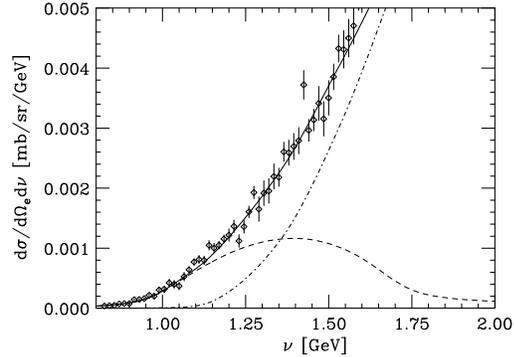}
\end{center}
\vspace*{-.5in}
\caption{\small
Comparison between the results of the approach of Ref. \protect\cite{Benhar91}
and the extrapolated nuclear matter cross section of Ref. \protect\cite{NMXS} at
$E_e = 3.6$ GeV and $\theta_{e^\prime}=30^\circ$. Dashed line: contribution of 
quasielastic scattering. Dash-dot line: contribution of inelastic channels.
Solid line: full calculation.  
}
\label{fig:4}
\vspace*{-.10in}
\end{figure}

In Figs. \ref{fig:3} and \ref{fig:4} the results of the approach of 
Ref. \cite{Benhar91} are compared to the cross section at beam energy 
$E_e = 3.6$ GeV and scattering angle $\theta_{e^\prime}=30^\circ$ 
(corresponding to $Q^2 \gsim 2$ GeV$^2$) obtained from the 
extrapolation of SLAC $(e,e^\prime)$ data to infinite $A$ \cite{NMXS}. 

Figure \ref{fig:3} clearly shows the dominance of FSI in the low energy loss
tail of the cross section, as well as the need of including of NN 
correlations to achieve a quantitative account of the data. 

The data displayed in Fig. \ref{fig:4} show that the transition from the 
quasi elastic to the inelastic regime, including resonant and nonresonant 
pion production as well as deep inelastic processes, is a smooth one, thus 
suggesting the possibility of a unified theoretical representation.
It appears that NMBT and the IA scheme provide a consistent and computationally 
viable approach, yielding a good description of the measured cross section over 
the whole $\nu$ range.

The energy loss spectra obtained applying the formalism discussed in the previous 
Sections to charged current neutrino-nucleus scattering exhibit qualitative 
features similar to those emerging from the analysis of electron-nucleus 
scattering \cite{Benhar05}.

\begin{figure}[hbt]
\begin{center}
\includegraphics[scale=0.35]{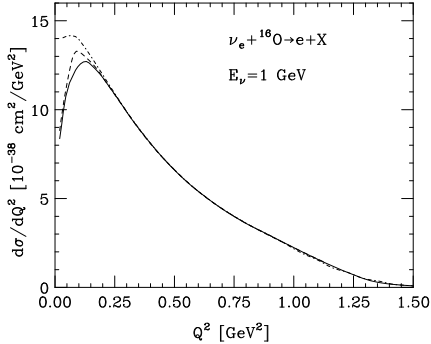}
\end{center}
\vspace*{-.5in}
\caption{\small
Differential cross section $d\sigma/dQ^2$
for neutrino energy $E_\nu= 1$ GeV. The dot-dash line shows the PWIA results,
while the solid and dashed lines have been obtained taking into acount the 
effect of Pauli blocking, with and without inclusion
of dynamical FSI, respectively.
}
\label{fig:5}
\vspace*{-.10in}
\end{figure}

The effect of Pauli blocking of the phase space available to the
knocked-out particle, while being hardly visible in Figs. \ref{fig:1}-\ref{fig:4},
is large in the $Q^2$ distributions at $Q^2 < 0.2$ GeV$^2$.
This feature is illustrated in Fig. \ref{fig:5}, showing
the calculated differential cross section $d\sigma/dQ^2$ of the process
$\nu_e~+~^{16}O~\rightarrow~e~+~X$, for neutrino 
energy $E_\nu= 1$ GeV \cite{Benhar05}.
The dashed and dot-dash lines correspond to the
PWIA results with and without inclusion of Pauli blocking, respectively. It
clearly appears that the effect of Fermi statistic in suppressing scattering
shows up at $Q^2 < 0.2$ GeV$^2$ and becomes very large at lower $Q^2$. The results of
the full calculation, in which dynamical FSI are also included, are displayed as 
a full line. 

Figure \ref{fig:5} suggests that Pauli blocking and FSI may explain
the deficit of the measured cross section at low $Q^2$ with respect to the
predictions of Monte Carlo simulations \cite{Ishida}.

\section{CONCLUSIONS}

The results discussed in this paper show that the approach based on NMBT 
provides quantitative {\it parameter free} predictions of the electroweak 
nuclear response in the impulse approximation regime, corresponding to beam 
energy larger than $\sim$ 1 GeV, relevant to many neutrino oscillation 
experiments. 

In the region of the quasi-elastic peak, theoretical results account
for the measured $^{16}O(e,e^\prime)$ cross sections at beam energies between
700 MeV and 1200 MeV and scattering angle 32$^\circ$ with an accuracy better
than 10 \% \cite{Benhar05}. Close agreement between theory and data is also  
found at larger energies, where inelastic processes dominate, with 
the only exception of the region of quasi-free $\Delta$ production, 
where theoretical predictions significantly underestimate the measured cross
sections. Although the disagreement is likely to be ascribable to 
uncertainties in the description of the nucleon structure functions at 
low $Q^2$, further studies are needed to clarify this issue.

The overall picture emerging from the comparison between theory and electron 
scattering data indicates that FSI are large and {\it do not} go away at large $Q^2$, 
as the total NN cross section, dominated by inelastic contributions, stays 
roughly constant over a broad energy range \cite{yscaling}. The main FSI effects 
 in both inclusive and 
semi-inclusive processes appear to be understood. 
The pivotal role played by NN correlation entails that a fully quantitative
treatment of FSI requires a realistic description of nuclear 
dynamics beyond the mean field approximation.
 
\section*{ACKNOWLEDGMENTS}

The results discussed in this paper have been obtained in collaboration 
with N.~Farina, H.~Nakamura, D.~Rohe, M.~Sakuda, R.~Seki and I.~Sick.
A number of illuminating discussions with A. Fabrocini, S. Fantoni 
and R. Schiavilla are also gratefully acknowledged.

\end{document}